\documentclass[rmp,twocolumn,floatfix]{revtex4}

\usepackage{graphicx}
\usepackage{bm}
\usepackage{url}

\begin{document}

\title{Statistical mechanics approach to the probability distribution of money}

\author{Victor M.~Yakovenko}

\affiliation{Department of Physics, University of
  Maryland, College Park, Maryland 20742-4111, USA}

\date{28 July 2010}


\begin{abstract}
This Chapter reviews statistical models for the probability distribution of money developed in the econophysics literature since the late 1990s.  In these models, economic transactions are modeled as random transfers of money between the agents in payment for goods and services.  Starting from the initially equal distribution of money, the system spontaneously develops a highly unequal distribution of money analogous to the Boltzmann-Gibbs distribution of energy in physics.  Boundary conditions are crucial for achieving a stationary distribution.  When debt is permitted, it destabilizes the system, unless some sort of limit is imposed on maximal debt. \\

\textsf{\normalsize ``Money, it's a gas.'' Pink Floyd, 
\textit{Dark Side of the Moon}}
\end{abstract}

\maketitle

\section{Introduction}
\label{Sec:money}

In this Chapter, the probability distribution of money in a system of economic agents is studied using methods of statistical physics.  This study originates from the interdisciplinary field known as econophysics \cite{Stanley-1996}, which applies mathematical methods of statistical physics to social, economical, and financial problems \cite{Stauffer-history}.  

One of the most puzzling social problems is the persistent wide range of economic inequality among the population in any society.  In statistical physics, it is very well known that identical (``equal'') molecules in a gas spontaneously develop a widely unequal distribution of energies as a result of random energy transfers in collisions between the molecules.  Using similar principles, this Chapter shows how a very unequal probability distribution of money among economic agents develops spontaneously as a result of money transfers between them.

The literature on social and economic inequality is enormous \cite{Kakwani-book}.
Many papers in the economic literature
\cite{Gibrat-1931,Kalecki-1945,Champernowne-1953} use a stochastic
process to describe dynamics of individual wealth or income and to
derive their probability distributions.  One might call this a
one-body approach, because wealth and income fluctuations are
considered independently for each economic agent.  Inspired by
Boltzmann's kinetic theory of collisions in gases, we
introduce an alternative, two-body approach, where agents perform
pairwise economic transactions and transfer money from one agent to
another.  We start with a simple pairwise-transfer model proposed by
\textcite{Dragulescu-2000}.  This model is the most closely related to
the traditional statistical mechanics, which we briefly review first.
Then we discuss other money-transfer models and further developments.  For a more detailed and systematic review of the progress in this field, see the recent review by \textcite{Yakovenko-2009}, as well as reviews by \textcite{Richmond-2006a,Richmond-2006b,Chatterjee-2007}
and a popular article by \textcite{Hayes-2002}.

Interestingly, the study of pairwise money transfer and the resulting
statistical distribution of money have virtually no counterpart in the modern economic literature.  Only the search theory of money \cite{Kiyotaki-1993} is somewhat related to it.  However, a probability distribution of money among the agents within the search-theoretical approach was only recently obtained numerically by the economist Miguel \textcite{Molico-2006}.

\section{The Boltzmann-Gibbs distribution of energy}
\label{Sec:BGphysics}

The fundamental law of equilibrium statistical mechanics is the
Boltzmann-Gibbs distribution.  It states that the probability
$P(\varepsilon)$ of finding a physical system or subsystem in a state
with the energy $\varepsilon$ is given by the exponential function
\begin{equation}
  P(\varepsilon)=c\,e^{-\varepsilon/T},
\label{Gibbs}
\end{equation}
where $T$ is the temperature, and $c$ is a normalizing constant
\cite{Wannier-book}. Here we set the Boltzmann constant $k_B$ to unity
by choosing the energy units for measuring the physical temperature
$T$.  Then, the expectation value of any physical variable $x$ can be
obtained as
\begin{equation}
  \langle x\rangle=\frac{\sum_k x_ke^{-\varepsilon_k/T}}
  {\sum_k e^{-\varepsilon_k/T}},
\label{expectation}
\end{equation}
where the sum is taken over all states of the system.  Temperature is
equal to the average energy per particle:
$T\sim\langle\varepsilon\rangle$, up to a numerical coefficient of the
order of 1.

Eq.\ (\ref{Gibbs}) can be derived in different ways
\cite{Wannier-book}.  All derivations involve the two main
ingredients: statistical character of the system and conservation of
energy $\varepsilon$.  One of the shortest derivations can be
summarized as follows.  Let us divide the system into two (generally
unequal) parts.  Then, the total energy is the sum of the parts:
$\varepsilon=\varepsilon_1+\varepsilon_2$, whereas the probability is
the product of probabilities:
$P(\varepsilon)=P(\varepsilon_1)\,P(\varepsilon_2)$.  The only
solution of these two equations is the exponential function
(\ref{Gibbs}).

A more sophisticated derivation, proposed by Boltzmann, uses
the concept of entropy.  Let us consider $N$ particles with the total
energy $E$.  Let us divide the energy axis into small intervals (bins)
of width $\Delta\varepsilon$ and count the number of particles $N_k$
having the energies from $\varepsilon_k$ to
$\varepsilon_k+\Delta\varepsilon$.  The ratio $N_k/N=P_k$ gives the
probability for a particle to have the energy $\varepsilon_k$.  Let us
now calculate the multiplicity $W$, which is the number of
permutations of the particles between different energy bins such that
the occupation numbers of the bins do not change.  This quantity is
given by the combinatorial formula in terms of the factorials
\begin{equation}
  W=\frac{N!}{N_1!\,N_2!\,N_3!\,\ldots}.
\label{multiplicity}
\end{equation}
The logarithm of multiplicity is called the entropy $S=\ln W$.  In the
limit of large numbers, the entropy per particle can be written in the
following form using the Stirling approximation for the factorials
\begin{equation}
  \frac{S}{N}=-\sum_k \frac{N_k}{N}\ln\left(\frac{N_k}{N}\right)
  =-\sum_k P_k\ln P_k.
\label{entropy}
\end{equation}
Now we would like to find what distribution of particles among
different energy states has the highest entropy, i.e.,\ the highest
multiplicity, provided the total energy of the system,
$E=\sum_kN_k\varepsilon_k$, has a fixed value.  Solution of this
problem can be easily obtained using the method of Lagrange
multipliers \cite{Wannier-book}, and the answer is given by the
exponential distribution (\ref{Gibbs}).

The same result can be also derived from the ergodic theory, which
says that the many-body system occupies all possible states of a given
total energy with equal probabilities.  Then it is straightforward to
show \cite{Lopez-Ruiz-2008} that the probability distribution of the
energy of an individual particle is given by Eq.\ (\ref{Gibbs}).

\section{Conservation of money}
\label{Sec:conservation}

The derivations outlined in Sec.\ \ref{Sec:BGphysics} are very general
and only use the statistical character of the system and the
conservation of energy.  So, one may expect that the exponential
Boltzmann-Gibbs distribution (\ref{Gibbs}) would apply to other
statistical systems with a conserved quantity.

The economy is a big statistical system with millions of participating
agents, so it is a promising target for applications of statistical
mechanics.  Is there a conserved quantity in the economy?
\textcite{Dragulescu-2000} argued that such a conserved quantity is
money $m$.  Indeed, the ordinary economic agents can only receive
money from and give money to other agents.  They are not permitted to
``manufacture'' money, e.g.,\ to print dollar bills.  Let us consider
an economic transaction between agents $i$ and $j$.  When the agent
$i$ pays money $\Delta m$ to the agent $j$ for some goods or services,
the money balances of the agents change as follows
\begin{eqnarray}
  && m_i\;\rightarrow\; m_i'=m_i-\Delta m,
\nonumber \\
  && m_j\;\rightarrow\; m_j'=m_j+\Delta m.
\label{transfer}
\end{eqnarray}
The total amount of money of the two agents before and after
transaction remains the same
\begin{equation}
  m_i+m_j=m_i'+m_j',
\label{conservation}
\end{equation}
i.e.,\ there is a local conservation law for money.  The rule
(\ref{transfer}) for the transfer of money is analogous to the
transfer of energy from one molecule to another in molecular
collisions in a gas, and Eq.\ (\ref{conservation}) is analogous to
conservation of energy in such collisions.  Conservative models of
this kind are also studied in some economic literature
\cite{Kiyotaki-1993,Molico-2006}.

We should emphasize that, in the model of \textcite{Dragulescu-2000}
[as in the economic models of \textcite{Kiyotaki-1993,Molico-2006}],
the transfer of money from one agent to another represents payment for
goods and services in a market economy.  However, the model of
\textcite{Dragulescu-2000} only keeps track of money flow, but does
not keep track of what goods and service are delivered.  One reason
for this is that many goods, e.g.,\ food and other supplies, and most
services, e.g.,\ getting a haircut or going to a movie, are not
tangible and disappear after consumption.  Because they are not
conserved, and also because they are measured in different physical
units, it is not very practical to keep track of them.  In contrast,
money is measured in the same unit (within a given country with a
single currency) and is conserved in local transactions
(\ref{conservation}), so it is straightforward to keep track of money
flow.  It is also important to realize that an increase in material
production does not produce an automatic increase in money supply.
The agents can grow apples on trees, but cannot grow money on trees.
Only a central bank has the monopoly of changing the monetary base
$M_b$ \cite{McConnell-book}.  (Debt and credit issues are discussed
separately in Sec.\ \ref{Sec:debt}.)

Enforcement of the local conservation law (\ref{conservation}) is the
key feature for successful functioning of money.  If the agents were
permitted to ``manufacture'' money, they would be printing money and
buying all goods for nothing, which would be a disaster.  The purpose of the conservation law is to ensure that an agent can buy goods and products from the society (the other agents) only if he or she contributes something useful to the society and receives money payment for these contributions.
Thus, money is an accounting device, and, indeed, all accounting systems are based on the conservation law (\ref{transfer}).  The physical
medium of money is not essential as long as the local
conservation law is enforced.  The days of gold standard are long
gone, so money today is truly the fiat money, declared to be money by
the central bank.  Money may be in the form of paper currency, but
today it is more often represented by digits on computerized 
accounts.  So, money is nothing but bits of information, and monetary system represents an informational layer of the economy.  Monetary system interacts with physical system (production and consumption of material goods), but the two layers cannot be transformed into each other because of their different nature.

Unlike, ordinary economic agents, a central bank or a central
government can inject money into the economy, thus changing the total
amount of money in the system.  This process is analogous to an influx
of energy into a system from external sources, e.g.,\ the Earth
receives energy from the Sun.  Dealing with these situations,
physicists start with an idealization of a closed system in thermal
equilibrium and then generalize to an open system subject to an energy
flux.  As long as the rate of money influx from central sources is
slow compared with relaxation processes in the economy and does not
cause hyperinflation, the system is in quasi-stationary statistical
equilibrium with slowly changing parameters.  This situation is
analogous to heating a kettle on a gas stove slowly, where the kettle
has a well-defined, but slowly increasing, temperature at any moment of
time.  A flux of money may be also produced by international transfers
across the boundaries of a country.  This process involves complicated
issues of multiple currencies in the world and their exchange rates
\cite{McCauley-2008}.  Here we consider an idealization of a closed economy
for a single country with a single currency.

Another potential problem with conservation of money is debt.  This
issue will be discussed in Sec.\ \ref{Sec:debt}.  As a
starting point, \textcite{Dragulescu-2000} considered simple models,
where debt is not permitted, which is also a common idealization in
some economic literature \cite{Kiyotaki-1993,Molico-2006}.  This means
that money balances of the agents cannot go below zero: $m_i\geq0$ for
all $i$.  Transaction (\ref{transfer}) takes place only when an agent
has enough money to pay the price: $m_i\geq\Delta m$, otherwise the
transaction does not take place.  If an agent spends all money, the
balance drops to zero $m_i=0$, so the agent cannot buy any goods from
other agents.  However, this agent can still receive money from other
agents for delivering goods or services to them.  In real life, money
balance dropping to zero is not at all unusual for people who live
from paycheck to paycheck.

Macroeconomic monetary policy issues, such as money supply and money
demand \cite{MonetaryEconomics-book}, are outside of the scope of this Chapter.  Our goal is to investigate the probability distribution of
money among economic agents.  For this purpose, it is appropriate to
make the simplifying macroeconomic idealizations, as described above,
in order to ensure overall stability of the system and existence of
statistical equilibrium.  The concept of ``equilibrium''
is a very common idealization in economic literature, even though the
real economies might never be in equilibrium.  Here we extend this
concept to a statistical equilibrium, which is characterized by a
stationary probability distribution of money $P(m)$, in contrast to a
mechanical equilibrium, where the ``forces'' of demand and supply
balance each other.

\section{The Boltzmann-Gibbs distribution of money}
\label{Sec:BGmoney}

Having recognized the principle of local money conservation,
\textcite{Dragulescu-2000} argued that the stationary distribution of
money $P(m)$ should be given by the exponential Boltzmann-Gibbs
function analogous to Eq.\ (\ref{Gibbs})
\begin{equation}
  P(m)=c\,e^{-m/T_m}.
\label{money}
\end{equation}
Here $c$ is a normalizing constant, and $T_m$ is the ``money
temperature'', which is equal to the average amount of money per
agent: $T=\langle m\rangle=M/N$, where $M$ is the total money, and $N$
is the number of agents.\footnote{Because debt is not permitted in
  this model, we have $M=M_b$, where $M_b$ is the monetary base
  \cite{McConnell-book}.}

\begin{figure}
\includegraphics[width=0.9\linewidth]{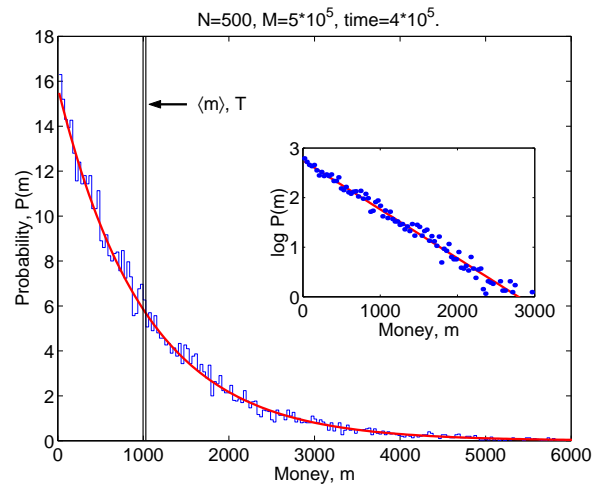}
\caption{\textit{Histogram and points:} Stationary probability
  distribution of money $P(m)$ obtained in additive random-exchange models.  \textit{Solid curves:} Fits to the exponential Boltzmann-Gibbs law
  (\ref{money}).  \textit{Vertical line:} The initial distribution of
  money.  From \textcite{Dragulescu-2000}.}
\label{Fig:money}
\end{figure}

To verify this conjecture, \textcite{Dragulescu-2000} performed
agent-based computer simulations of money transfers between agents.
Initially all agents were given the same amount of money, say, \$1000.
Then, a pair of agents $(i,j)$ was randomly selected, the amount
$\Delta m$ was transferred from one agent to another, and the process
was repeated many times.  Time evolution of the probability
distribution of money $P(m)$ is illustrated in computer animation videos
by \textcite{Chen-2007} and by \textcite{Wright-2007}.  After a
transitory period, money distribution converges to the stationary form
shown in Fig.\ \ref{Fig:money}.  As expected, the distribution is 
well fitted by the exponential function (\ref{money}).

Several different rules for $\Delta m$ were considered by
\textcite{Dragulescu-2000}.  In one model, the transferred amount was
fixed to a constant $\Delta m=\$1$.  Economically, it means that all
agents were selling their products for the same price $\Delta m=\$1$.
Computer animation \cite{Chen-2007} shows that the initial
distribution of money first broadens to a symmetric Gaussian curve,
characteristic for a diffusion process.  Then, the distribution starts
to pile up around the $m=0$ state, which acts as the impenetrable
boundary, because of the imposed condition $m\geq0$.  As a result,
$P(m)$ becomes skewed (asymmetric) and eventually reaches the
stationary exponential shape, as shown in Fig.\ \ref{Fig:money}.  The
boundary at $m=0$ is analogous to the ground-state energy in
statistical physics.  Without this boundary condition, the probability
distribution of money would not reach a stationary state.  Computer
animations \cite{Chen-2007,Wright-2007} also show how the entropy of
money distribution, defined as $S/N=-\sum_k P(m_k)\ln P(m_k)$, grows
from the initial value $S=0$, where all agents have the same money, to
the maximal value at the statistical equilibrium.

While the model with $\Delta m=1$ is very simple and instructive, it
is not realistic, because all prices are taken to be the same.
In another model considered by \textcite{Dragulescu-2000}, $\Delta m$
in each transaction is taken to be a random fraction of the average
amount of money per agent, i.e.,\ $\Delta m=\nu(M/N)$, where $\nu$ is
a uniformly distributed random number between 0 and 1.  The random
distribution of $\Delta m$ is supposed to represent the wide variety
of prices for different products in the real economy.  Computer
simulation of this model produces the same stationary
distribution (\ref{money}).  \textcite{Dragulescu-2000}
also considered a model with firms, which hire agents to produce and sell products.   This process results in a many-body transfer of money, as opposed to pairwise transfer discussed above.  Computer simulation of this model also generates the same exponential distribution (\ref{money}).

These ideas were further developed by 
\textcite{Scalas-2006,Garibaldi-2007}.  The Boltzmann distribution was
independently applied to social sciences by the physicist J\"urgen
\textcite{Mimkes-2000,Mimkes-2005} using the Lagrange principle of
maximization with constraints.  The exponential distribution of money
was also found by the economist Martin \textcite{Shubik-1999} using a
Markov chain approach to strategic market games.  A long time ago,
Benoit \textcite[p 83]{Mandelbrot-1960} observed:
\begin{quote}
  ``There is a great temptation to consider the exchanges of money
  which occur in economic interaction as analogous to the exchanges of
  energy which occur in physical shocks between gas molecules.''
\end{quote}
He realized that this process should result in the exponential
distribution, by analogy with the barometric distribution of density
in the atmosphere.  However, he discarded this idea, because it does
not produce the Pareto power law \cite{Pareto-book}, and proceeded to study the stable L\'evy distributions.  Ironically, the actual economic data \cite{Yakovenko-2009} do show the exponential distribution for the majority of the population.  Moreover, the data have a finite variance, so the stable L\'evy
distributions are not applicable because of their infinite variance.

\section{Proportional money transfers and saving propensity}
\label{Sec:saving}

\begin{figure}
\includegraphics[width=0.9\linewidth]{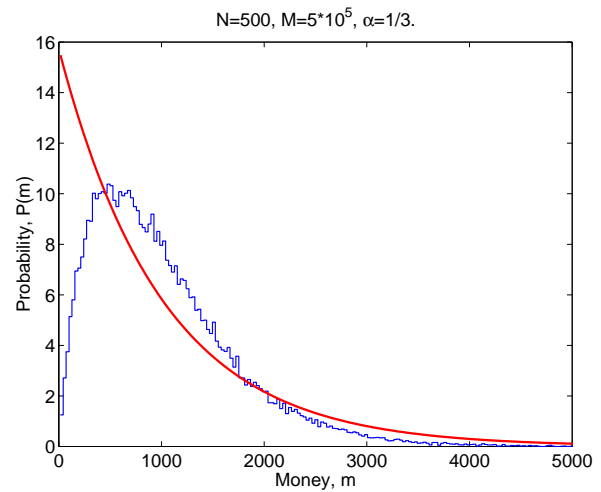}
\caption{\textit{Histogram:} Stationary probability distribution of
  money in the multiplicative random-exchange model
  (\ref{proportional}) for $\gamma=1/3$.  \textit{Solid curve:} The
  exponential Boltzmann-Gibbs law.  From
  \textcite{Dragulescu-2000}.}
\label{Fig:Redner}
\end{figure}

In the models of money transfer discussed in Sec.\ \ref{Sec:BGmoney},
the transferred amount $\Delta m$ is typically independent of the
money balances of the agents involved.  A different model was introduced
earlier by the physicists \textcite{Ispolatov-1998}
and called the multiplicative asset exchange model.  This model also
satisfies the conservation law, but the transferred amount of money is
a fixed fraction $\gamma$ of the payer's money in
Eq.\ (\ref{transfer}):
\begin{equation}
  \Delta m=\gamma m_i.
\label{proportional}
\end{equation}
The stationary distribution of money in this model, compared in
Fig.\ \ref{Fig:Redner} with an exponential function, is similar, albeit
not exactly equal, to the Gamma distribution:
\begin{equation}
  P(m)=c\,m^\beta\,e^{-m/T}.
\label{Gamma}
\end{equation}
Eq.\ (\ref{Gamma}) differs from Eq.\ (\ref{money}) by the power-law
prefactor $m^\beta$.  For $\beta>0$, the
population with low money balances is reduced, and $P(0)=0$, as shown
in Fig.\ \ref{Fig:Redner}.

Essentially the same model \cite{Lux-2005}, called the inequality
process, has been introduced and studied much earlier by the
sociologist John
\textcite{Angle-1986,Angle-1992,Angle-1993,Angle-1996,Angle-2002,Angle-2006}.  \textcite{Angle-1986} associated the proportionality rule (\ref{proportional})
with the surplus theory of social stratification
\cite{Engels-book}, which argues that inequality in human society
develops when people can produce more than necessary for minimal
subsistence.  This additional wealth (surplus) can be transferred from
original producers to other people, thus generating inequality.  Angle found a Gamma-like distribution (\ref{Gamma}) in numerical simulations of his models.    Independently, the economist Miguel \textcite{Molico-2006} studied conservative exchange models (\ref{transfer}) where agents bargain over prices in their transactions.  He also found a stationary Gamma-like distribution of money in numerical simulations.  

Another model with an element of proportionality was proposed by
\textcite{Chakraborti-2000}.  In this model, the
agents set aside (save) some fraction of their money $\lambda m_i$,
whereas the rest of their money balance $(1-\lambda)m_i$ becomes
available for random exchanges.  Thus, the rule of exchange
(\ref{transfer}) becomes
\begin{eqnarray}
  && m_i'=\lambda m_i + \xi(1-\lambda)(m_i+m_j),
\nonumber \\
  && m_j'=\lambda m_j + (1-\xi)(1-\lambda)(m_i+m_j).
\label{saving}
\end{eqnarray}
Here the coefficient $\lambda$ is called the saving propensity, and
the random variable $\xi$ is uniformly distributed between 0 and 1.
Computer simulations by \textcite{Chakraborti-2000} of the model (\ref{saving})
found a stationary distribution close to the Gamma distribution
(\ref{Gamma}).  With $\lambda\neq0$, agents always keep some money, so their balances never drop to zero, thus $P(0)=0$.

In the subsequent papers by the Kolkata school and related papers, the case of random saving propensity was studied.  In these models, the agents are
assigned random parameters $\lambda$ drawn from a uniform distribution
between 0 and 1 \cite{Chatterjee-2004}.  It was found that this model
produces a power-law tail $P(m)\propto1/m^2$ at high $m$.  The reasons
for stability of this law were understood using the Boltzmann kinetic
equation \cite{Das-2005,Chatterjee-2005,Repetowicz-2005}, but most
elegantly in the mean-field theory
\cite{Mohanty-2006,Bhattacharyya-2007,Chatterjee-2007}.  The fat tail
originates from the agents whose saving propensity is close to 1, who
hoard money and do not give it back
\cite{Patriarca-2005,Patriarca-2006}.  A more rigorous mathematical
treatment of the problem was given by
\textcite{During-2008,Matthes-2008,During-2007}.  

As a further extension, \textcite{Dragulescu-2000} considered a model with taxation, which also has an element of proportionality.  The Gamma distribution was
also studied for conservative models within a simple Boltzmann
approach by \textcite{Ferrero-2004,Ferrero-2005} and, using more complicated rules
of exchange motivated by political economy, by
\textcite{Scafetta-2004a,Scafetta-2004b}.  Another extension of these studies includes not only money transfers, but also transfers of a commodity, for which money is paid 
\cite{Chakraborti-2001,Chatterjee-2006,Ausloos-2007,Silver-2002,Lux-2009}.  For a more detailed review of these models, see \textcite{Yakovenko-2009}.

The stationary distribution of money (\ref{Gamma}) is different from the simple exponential formula (\ref{money}).
The origin of this difference can be understood from the Boltzmann
kinetic equation \cite{Wannier-book,Kinetics-book}.  This equation
describes time evolution of the distribution function $P(m)$ due to
pairwise interactions:
\begin{eqnarray}
  &&\frac{dP(m)}{dt}=\int\!\!\!\!\int\{
    -f_{[m,m']\to[m-\Delta,m'+\Delta]}P(m)P(m')
\label{Boltzmann}  \\
  &&+f_{[m-\Delta,m'+\Delta]\to[m,m']}
  P(m-\Delta)P(m'+\Delta)\}\,dm'\,d\Delta.
\nonumber
\end{eqnarray}
Here $f_{[m,m']\to[m-\Delta,m'+\Delta]}$ is the probability of
transferring money $\Delta$ from an agent with money $m$ to an agent
with money $m'$ per unit time.  This probability, multiplied by the
occupation numbers $P(m)$ and $P(m')$, gives the rate of transitions
from the state $[m,m']$ to the state $[m-\Delta,m'+\Delta]$.  The
first term in Eq.\ (\ref{Boltzmann}) gives the depopulation rate of
the state $m$.  The second term in Eq.\ (\ref{Boltzmann}) describes
the reversed process, where the occupation number $P(m)$ increases.
When the two terms are equal, the direct and reversed transitions
cancel each other statistically, and the probability distribution is
stationary: $dP(m)/dt=0$.  This is the principle of detailed balance.

In physics, the fundamental microscopic equations of motion obey the
time-reversal symmetry.  This means that the probabilities of the
direct and reversed processes are exactly equal:
\begin{eqnarray}
  f_{[m,m']\to[m-\Delta,m'+\Delta]}=f_{[m-\Delta,m'+\Delta]\to[m,m']}.
\label{reversal}
\end{eqnarray}
When Eq.\ (\ref{reversal}) is satisfied, the detailed balance
condition for Eq.\ (\ref{Boltzmann}) reduces to the equation
$P(m)P(m')=P(m-\Delta)P(m'+\Delta)$, because the factors $f$ cancels
out.  The only solution of this equation is the exponential function
$P(m)=c\exp(-m/T_m)$, so the Boltzmann-Gibbs distribution is the
stationary solution of the Boltzmann kinetic equation
(\ref{Boltzmann}).  Notice that the transition probabilities
(\ref{reversal}) are determined by the dynamical rules of the model,
but the equilibrium Boltzmann-Gibbs distribution does not depend on
the dynamical rules at all.  This is the origin of the universality of
the Boltzmann-Gibbs distribution.  We see that it is possible to find
the stationary distribution without knowing details of the dynamical
rules (which are rarely known very well), as long as the symmetry
condition (\ref{reversal}) is satisfied.

The models considered in Sec.\ \ref{Sec:BGmoney} have the
time-reversal symmetry.  The model with the fixed money transfer
$\Delta$ has equal probabilities (\ref{reversal}) of transferring
money from an agent with the balance $m$ to an agent with the balance
$m'$ and vice versa.  This is also true when $\Delta$ is random, as
long as the probability distribution of $\Delta$ is independent of $m$
and $m'$.  Thus, the stationary distribution $P(m)$ is always
exponential in these models.  On the other hand, in the model
(\ref{proportional}), the time-reversal symmetry is broken.  Indeed,
when an agent $i$ gives a fixed fraction $\gamma$ of his money $m_i$
to an agent with balance $m_j$, their balances become $(1-\gamma)m_i$
and $m_j+\gamma m_i$.  If we try to reverse this process and appoint
the agent $j$ to be the payer and to give the fraction $\gamma$ of her
money, $\gamma(m_j+\gamma m_i)$, to the agent $i$, the system does not
return to the original configuration $[m_i,m_j]$.  As emphasized by
\textcite{Angle-2006}, the payer pays a deterministic fraction of his
money, but the receiver receives a random amount from a random agent,
so their roles are not interchangeable.  Because the proportional rule
typically violates the time-reversal symmetry, the stationary
distribution $P(m)$ in multiplicative models is not exponential.

These examples show that the Boltzmann-Gibbs distribution does not
necessarily hold for any conservative model.  However, it is universal
in a limited sense for a broad class of models that have
time-reversal symmetry.  In the absence of detailed knowledge of real microscopic dynamics of economic exchanges, the semiuniversal Boltzmann-Gibbs distribution
(\ref{money}) is a natural starting point.  Moreover, the assumption
of \textcite{Dragulescu-2000} that agents pay the same prices $\Delta
m$ for the same products, independent of their money balances $m$,
seems very appropriate for the modern anonymous economy, especially
for purchases over the Internet.  There is no particular empirical
evidence for the proportional rules (\ref{proportional}) or
(\ref{saving}).  By further modifying the rules of money transfer and introducing more parameters in the models, it is possible to obtain even more complicated
distributions \cite{Scafetta-2007}.  However, parsimony is the virtue of a good mathematical model, not the abundance of additional assumptions and parameters, whose correspondence to reality is hard to verify.

\section{Models with debt}
\label{Sec:debt}

Now let us discuss how the results change when debt is
permitted.\footnote{The ideas presented here are quite similar to
  those by \textcite{Soddy-book}.  Frederick Soddy, the Nobel Prize winner in chemistry for his work on radioactivity, argued that the real wealth is derived
from the energy use in transforming raw materials into goods and
services, and not from monetary transactions.  He also warned about
dangers of excessive debt and related ``virtual wealth'' resulting in the Great Depression.}  From the standpoint of individual
economic agents, debt may be considered as negative money.  When an
agent borrows money from a bank (considered here as a big reservoir of
money),\footnote{Here we treat the bank as being outside of the system
  consisting of ordinary agents, because we are interested in money
  distribution among these agents.  The debt of agents is an asset for
  the bank, and deposits of cash into the bank are liabilities of the
  bank \cite{McConnell-book}.  We do not go into these details in
  order to keep our presentation simple.} the cash balance of the agent (positive
money) increases, but the agent also acquires a debt obligation
(negative money), so the total balance (net worth) of the agent
remains the same.  Thus, the act of borrowing money still satisfies a
generalized conservation law of the total money (net worth), which is
now defined as the algebraic sum of positive (cash $M$) and negative
(debt $D$) contributions: $M-D=M_b$, where $M_b$ is the original amount of money in the system, the monetary base \cite{McConnell-book}.  After spending some cash in
pairwise transactions (\ref{transfer}), the agent still has the debt
obligation (negative money), so the total money balance $m_i$ of the
agent (net worth) becomes negative.  We see that the boundary
condition $m_i\geq0$, discussed in Sec.\ \ref{Sec:conservation}, does
not apply when debt is permitted, so $m=0$ is not the ground state any
more.  The consequence of permitting debt is not a violation of the
conservation law (which is still preserved in the generalized form for
net worth), but a modification of the boundary condition by permitting
agents to have negative balances $m_i<0$ of net worth.  A more
detailed discussion of positive and negative money and the
book-keeping accounting from the econophysics point of view was
presented by the physicist Dieter \textcite{Braun-2001} and
\textcite{Fischer-2003a,Fischer-2003b}.

Now we can repeat the simulation described in Sec.\ \ref{Sec:BGmoney}
without the boundary condition $m\geq0$ by allowing agents to go into
debt.  When an agent needs to buy a product at a price $\Delta m$
exceeding his money balance $m_i$, the agent is now permitted to
borrow the difference from a bank and, thus, to buy the product.  As a
result of this transaction, the new balance of the agent becomes
negative: $m_i'=m_i-\Delta m<0$.  Notice that the local conservation
law (\ref{transfer}) and (\ref{conservation}) is still satisfied, but
it involves negative values of $m$.  If the simulation is continued
further without any restrictions on the debt of the agents, the
probability distribution of money $P(m)$ never stabilizes, and the
system never reaches a stationary state.  As time goes on, $P(m)$
keeps spreading in a Gaussian manner unlimitedly toward $m=+\infty$
and $m=-\infty$.  Because of the generalized conservation law
discussed above, the first moment $\langle m\rangle=M_b/N$ of the
algebraically defined money $m$ remains constant.  It means that some
agents become richer with positive balances $m>0$ at the expense of
other agents going further into debt with negative balances $m<0$, so
that $M=M_b+D$.

Common sense, as well as the experience with the current financial
crisis, tells us that an economic system cannot be stable if unlimited
debt is permitted.\footnote{In qualitative agreement with the 
conclusions by \textcite{McCauley-2008}.}  In this case, agents can
buy any goods without producing anything in exchange by simply going
into unlimited debt.  Arguably, the current financial crisis was
caused by the enormous debt accumulation in the system, triggered by
subprime mortgages and financial derivatives based on them.  A widely
expressed opinion is that the current crisis is not the problem of
liquidity, i.e.,\ a temporary difficulty in cash flow, but the problem
of insolvency, i.e.,\ the inherent inability of many participants to pay
back their debts.

\begin{figure}
\includegraphics[width=0.9\linewidth]{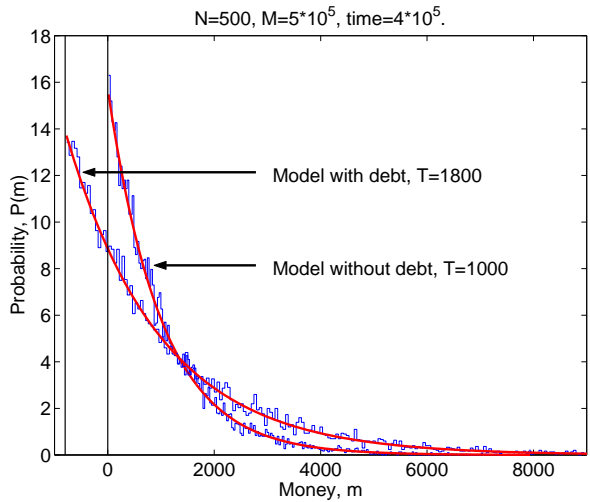}
\caption{\textit{Histograms:} Stationary distributions of money with
  and without debt.  The debt is limited to $m_d=800$. \textit{Solid
  curves:} Fits to the Boltzmann-Gibbs laws with the ``money
  temperatures'' $T_m=1800$ and $T_m=1000$.  From
  \textcite{Dragulescu-2000}.}
\label{Fig:debt}
\end{figure}

Detailed discussion of the current economic situation is not a subject
of this paper.  Going back to the idealized model of money transfers,
one would need to impose some sort of modified boundary conditions in
order to prevent unlimited growth of debt and to ensure overall
stability of the system.  \textcite{Dragulescu-2000} considered a
simple model where the maximal debt of each agent is limited to a
certain amount $m_d$.  This means that the boundary condition
$m_i\geq0$ is now replaced by the condition $m_i\geq-m_d$ for all
agents $i$.  Setting interest rates on borrowed money to be zero for
simplicity, \textcite{Dragulescu-2000} performed computer simulations
of the models described in Sec.\ \ref{Sec:BGmoney} with the new
boundary condition.  The results are shown in Fig.\ \ref{Fig:debt}.
Not surprisingly, the stationary money distribution again has the
exponential shape, but now with the new boundary condition at $m=-m_d$
and the higher money temperature $T_d=m_d+M_b/N$.  By allowing agents
to go into debt up to $m_d$, we effectively increase the amount of
money available to each agent by $m_d$.  So, the money temperature,
which is equal to the average amount of effectively available money
per agent, increases correspondingly.

\begin{figure}
\includegraphics[width=\linewidth]{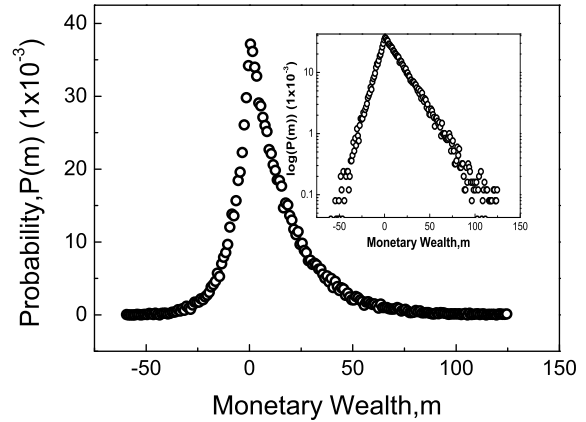}
\caption{The stationary distribution of money for the required reserve
  ratio $R=0.8$.  The distribution is exponential for both positive and
  negative money with different ``temperatures'' $T_+$ and $T_-$, as
  illustrated by the inset on log-linear scale.  From
  \textcite{Xi-Ding-Wang-2005}.}
\label{Fig:reserve}
\end{figure}

\textcite{Xi-Ding-Wang-2005} considered another, more realistic
boundary condition, where a constraint is imposed not on the
individual debt of each agent, but on the total debt of all agents in
the system.  This is accomplished via the required reserve ratio $R$,
which is briefly explained below \cite{McConnell-book}.  Banks are
required by law to set aside a fraction $R$ of the money deposited
into bank accounts, whereas the remaining fraction $1-R$ can be loaned
further.  If the initial amount of money in the system (the money
base) is $M_b$, then, with repeated loans and borrowing, the total
amount of positive money available to the agents increases to
$M=M_b/R$, where the factor $1/R$ is called the money multiplier
\cite{McConnell-book}.  This is how ``banks create money''.  Where
does this extra money come from?  It comes from the increase in the
total debt in the system.  The maximal total debt is given by
$D=M_b/R-M_b$ and is limited by the factor $R$.  When the debt is
maximal, the total amounts of positive, $M_b/R$, and negative,
$M_b(1-R)/R$, money circulate among the agents in the system, so there
are two constraints in the model considered by
\textcite{Xi-Ding-Wang-2005}.  Thus, we expect to see the exponential
distributions of positive and negative money characterized by two
different temperatures: $T_+=M_b/RN$ and $T_-=M_b(1-R)/RN$.  This is
exactly what was found in computer simulations by
\textcite{Xi-Ding-Wang-2005}, as shown in Fig.\ \ref{Fig:reserve}.
Similar two-sided distributions were also found by
\textcite{Fischer-2003a}.

However, in the real economy, the reserve requirement is not effective in
stabilizing total debt in the system, because it applies only to
deposits from general public, but not from corporations.  Moreover, there are alternative instruments of debt, including derivatives and various unregulated ``financial innovations''.  As a result, the total debt is not limited in practice
and can potentially reach catastrophic proportions.  Here we briefly
discuss several models with non-stationary debt.  Thus far, we did not
consider the interest rates.  \textcite{Dragulescu-2000} studied a
simple model with different interest rates for deposits into and loans
from a bank.  Computer simulations found that money distribution among
the agents is still exponential, but the money temperature slowly
changes in time.  Depending on the choice of parameters, the total
amount of money in circulation either increases or decreases in time.
Interest amplifies destabilizing effect of debt, because positive balances become even more positive and negative even more negative due to accruement of interest.
A more sophisticated macroeconomic model studied by the economist
Steve \textcite{Keen-1995,Keen-2000} exhibits debt-induced breakdown, where all economic activity stops under the burden of heavy debt and cannot be restarted without a ``debt moratorium''.  The interest rates were fixed in these models
and not adjusted self-consistently.  \textcite{Cockshott-2008}
proposed a mechanism, where the interest rates are set to cover
probabilistic withdrawals of deposits from a bank.  In an agent-based
simulation of the model, \textcite{Cockshott-2008} found that money
supply first increases up to a certain limit, and then the economy
experiences a spectacular crash under the weight of accumulated debt.

In the absence of a nominal limit on maximal debt, bankruptcy provides a mechanism for debt stabilization.  When the debt of an agent becomes too large, the agent will not be able to borrow any more money and will not be able to pay the debt back, so he or she will have to declare bankruptcy.  Bankruptcy erases the debt of the agent (the negative money) and resets the balance to zero.  However, somebody else (a bank or a lender) counted this debt as a positive asset, which also becomes erased.  In the language of physics, creation of debt is analogous to particle-antiparticle generation (creation of positive and negative money), whereas cancellation of debt by repayment or by bankruptcy corresponds to particle-antiparticle annihilation (annihilation of positive and negative money).  The former process dominates during economic bubbles (booms) and represents monetary expansion, whereas the latter dominates during the subsequent recessions (busts) and represents monetary contraction.  Bankruptcy is the crucial mechanism for stabilizing money distribution, but it is often overlooked by the economists.  Interest rates are meaningless without a mechanism specifying when bankruptcy is triggered. 

After lending money out, the lender has the burden of collecting the debt from the debtor.  Thus, the act of lending creates a persistent link (a string) between the lender and the borrower, as emphasized in the Chapter by Heiner Ganssmann in this Volume.  This is in contrast to payments by positive money for goods and services, which are final and do not leave any persistent link between the agents after the transaction.  Invention of the infamous collateralized debt obligations (CDO) obscured connections between lenders and borrowers by randomizing and anonymizing their pools.  It destabilized the system by inviting unsustainable debt and made bankruptcy proceedings extremely difficult because of the scrambled identities of lenders and borrowers.

As argued above, boundary conditions are crucial for stabilizing money distribution.  Typically, a lower bound is imposed, but not an upper bound (in a capitalist, as opposed to a socialist, society).\footnote{If an upper limit is imposed instead of a lower limit, the money temperature in Eq.~(\ref{money}) becomes negative $T_m<0$, so the slope in Fig.~\ref{Fig:money} changes to $dP/dm>0$, which is known in physics as the inverse population.  The case with both upper and lower limits was studied by \textcite{Dragulescu-2000}.}  This asymmetry is very important for stability of a monetary system.  Numerous attempts were made to create alternative community money from scratch and most of them failed.  In such a system, when an agent provides goods or services to another agent, their accounts are credited with positive and negative tokens, as in Eq.\ (\ref{transfer}).  However, because the initial global money balance is zero in this case, the probability distribution of money $P(m)$ is symmetric with respect to positive and negative $m$.  Unless a boundary condition is imposed on the lower side, $P(m)$ will never stabilize.  Some agents will accumulate unlimited negative balance by consuming goods and services and not contributing anything in return, thus undermining the system.  In contrast, when a central government creates positive money by fiat and forces its usage by demanding that taxes are paid with this money, it creates a viable monetary system, as discussed in the Chapter by Randall Wray in this Volume.  Thus, taxation is an essential ingredient for vitality of a monetary system.

\section{Conclusions and perspectives}
\label{Sec:conclusions}

In this Chapter, we have demonstrated that random transfers of money in economic transactions between otherwise equal economic agents produce a broad and highly unequal probability distribution of money among the agents.  In additive models, the probability distribution of money is exponential and similar to the Boltzmann-Gibbs distribution of energy in statistical physics.  Multiplicative models produce a Gamma-like distribution and a power-law tail for random saving propensity.  Local conservation of money in transactions between agents is crucial for the accounting function of money.  Ordinary economic agents can only receive and give money, but cannot produce it.  Boundary conditions are necessary in order to achieve a stable probability distribution of money.  Without debt, zero money balance is the boundary.  When debt is permitted, some sort of restriction on the negative money balances must be imposed, either by limiting individual or collective debt or by setting up conditions for bankruptcy.  When debt is unlimited, the system is unstable and does not have a stationary state. 

It would be very interesting to compare these theoretical conclusions with empirical data on money distribution.  Unfortunately, it is very difficult to obtain such data.  The probability distribution of balances on deposit accounts in a big enough bank would be a reasonable approximation for money distribution among the population. However, such data are not publicly available.  In contrast, plenty of data are available on income distribution from the tax agencies.  Quantitative analysis of such data for the USA \cite{Yakovenko-2009} shows that the population consists of two distinct social classes.  Income distribution follows the exponential law for the lower class (about 97\% of population) and the power law for the upper class (about 3\% of population).   Although social classes have been known since Karl Marx, it is interesting that they can be recognized by fitting the empirical data with simple mathematical functions.  Using sophisticated models of interacting economic agents, the computer scientist Ian \textcite{Wright-2005,Wright-2009} demonstrated emergence of two classes in agent-based simulations of initially equal agents.  This work has been further developed in the book by \textcite{Cockshott-2009}, integrating economics,
computer science, and physics.

Nowadays, money is typically represented by data bits on computer accounts, which constitute the informational layer of the economy.  In contrast, material standards of living are determined by the physical layer of the economy.  In the modern society, physical standards of living are largely determined by the level of energy consumption and are widely different around the globe.   \textcite{Banerjee-2010} found that the probability distribution of energy consumption per capita around the world approximately follows the exponential law.  So, it is likely that the energy consumption inequality is governed by the same principles as the money inequality.  The energy/ecology and financial/economic crises are the biggest challenges faced by the mankind today.  There is an urgent need to find ways for a manageable and realistic transition from the current breakneck growth-oriented economy, powered by ever-expanding use of fossil energy fuels, to a stable and sustainable society, based on renewable energy and balance with the Nature.  Undoubtedly, both money and energy will be the key factors shaping up the future of human civilization.


\end{document}